\documentclass{article}
\usepackage{frascatiphys}
\usepackage{graphicx}
\usepackage{amsmath}
\usepackage{txfonts}

\usepackage[backend=biber,style=nature]{biblatex}
\begin{document}
\title{ 
CONSTRAINTS ON DARK MATTER FROM REIONIZATION
}
\author{
Marco Castellano   \\
\textit{INAF - Osservatorio Astronomico di Roma,}\\ \textit{via di Frascati 33, 00078 Monte Porzio Catone (RM), Italy} \\
Nicola Menci  \\
\textit{INAF - Osservatorio Astronomico di Roma,}\\\textit{ via di Frascati 33, 00078 Monte Porzio Catone (RM), Italy} \\
Massimiliano Romanello \\
\textit{INAF - Osservatorio Astronomico di Roma,}\\ \textit{via di Frascati 33, 00078 Monte Porzio Catone (RM), Italy;}\\ \textit{Dipartimento di Fisica e Astronomia - Alma Mater Studiorum Università di Bologna,}\\ \textit{via Piero Gobetti 93/2, I-40129 Bologna, Italy;}\\ \textit{INAF - Osservatorio di Astrofisica e Scienza dello Spazio di Bologna,}\\ \textit{via Piero Gobetti 93/3, I-40129 Bologna, Italy} \\
}
\maketitle
\baselineskip=11.6pt
\begin{abstract}
This conference proceedings paper provides a short summary of the constraints presented in Menci et al. 2016, 2017 \cite{Menci,Menci2017} on the mass of thermal WDM candidates, and of the results presented in Romanello et al. 2021 \cite{Romanello21} on how Reionization scenarios are affected by early galaxy formation in WDM cosmologies.
The abundance of galaxies in the epoch of reionization ($z>$6) is dependent on fundamental cosmological parameters, most importantly on the properties of dark matter, such that it can be used as a powerful cosmological probe. Here we show how the number density of primordial galaxies allows to constrain the mass of thermal WDM candidates, and the constraints that will be made possible by future JWST observations. We then investigate how the Reionization process is affected by early galaxy formation in different cosmological scenarios. We use a semi-analytic model with suppressed initial power spectra to obtain the UV Luminosity Function in thermal Warm Dark Matter and sterile neutrino cosmologies. For each cosmology, we find an upper limit to fixed $f_{esc}$, which guarantees the completion of the process at $z<6.7$. 
\end{abstract}
\baselineskip=14pt

\section{Introduction}

The Epoch of Reionization (EoR) marked a fundamental phase transition in the history of the Universe, during which the Intergalactic Medium (IGM) became transparent to UV photons. The most recent observations indicate a late-Reionization scenario~\cite{Hoag19, Mason19, Planck18}, with the end of the EoR at $z \approx 6$, but the exact contribution from different ionizing sources and the exact timeline and topology of Reionization are still unknown.

From a theoretical point-of-view, cosmic Reionization depends on non-linear and non-local phenomena, in which the physics of galaxy formation couples with the physics of gravity and radiation transport. The first process is determined by both baryonic physics and poorly known feedback effects, but also by the initial power spectrum of density fluctuations: in fact, dark matter produces the potential wells in which baryonic perturbation undergo an accelerated growth. Therefore, the study of Reionization is strongly related to the comprehension of cosmological framework in which cosmic structures form and grow. 

The currently most acknowledged cosmological model is the $\Lambda$CDM model. It is based on the contribution of the cosmological constant $\Lambda$ ($\approx$$69 \%$) and Cold Dark Matter ($\approx$$26\%$) and provides a coherent large-scale description of the Universe with respect to the available data. The $\Lambda$CDM model postulates the existence of Dark Matter in a ``cold'' version, i.e., composed by Weakly Interacting Massive Particles (WIMPs) with $m_X>0.1$ GeV or condensates of light axions, with $m_X\approx10^{-5}$--$10^{-1}$ eV.

However, there are some possible tensions related to observations at galactic and sub-galactic scales, of the order of kpc. 
Furthermore, the lack of detection of CDM candidates has suggested the possibility to investigate on alternative cosmological scenarios, based on the existence of Warm Dark Matter particles, with mass of the order of keV. While in the $\Lambda$CDM model, due to the high mass particles, all the cosmological density perturbations can become gravitationally unstable, in a WDM scenario, depending on the value of $m_X$, only perturbations above the kpc scale can collapse, producing shallower density profiles and a smaller number of low-mass halos. This, in the context of the hierarchical growth of the cosmic structures, implies a reduction in the number of faint galaxies and a delay in their formation~\cite{Menci,Dayal15}.

In WDM cosmologies, the simplest approach is to consider particles that behave as ``thermal relics'', resulting from the freeze-out of DM species initially in thermal equilibrium with the early Universe, e.g. ~\cite{Viel,Corasaniti}. 

A possible alternative is offered by sterile neutrinos (SN) or right-handed neutrinos, which are particles predicted in the context of Standard Model extensions. Since they are produced out-of-equilibrium, from the oscillations of active neutrinos, they are characterized by a non-thermal power spectrum, which depends both on mass and on $\sin(2\theta)$, where $\theta$ is the mixing angle~\cite{Merle}.

In the present paper we first summarise the results presented in Menci et al. 2016,2017 \cite{Menci,Menci2017} where stringent constraints on DM models with suppressed power spectra by have been derived by comparing the maximum number density of DM halos $\overline{\phi}$ expected at redshift $z=6$ to the observed number density ${\phi}_{obs}$ of galaxies at the same redshift in the HFF. We then summarise the results presented in Romanello et al. 2021 \cite{Romanello21} on how Reionization scenarios are affected by early galaxy formation in WDM cosmologies. We have used the theoretical model by Menci et al. (2018)~\cite{Menci18}, where the collapse history of dark matter halos is modelled through the Extended Press-Schechter (EPS) formalism and baryonic processes taking place in each halo are included through physically motivated analytical recipes.

\section{Method}

\subsection{Warm Dark Matter thermal relics}
 
The simplest alternative to CDM is provided by Warm Dark Matter models assuming DM to be the result from the freeze-out of particles  with mass in the keV range initially in thermal equilibrium in the early Universe. In these models, the population of low-mass galaxies is characterized by lower abundances and shallower central density profiles compared to Cold Dark Matter (CDM) due to the dissipation of small-scale density perturbations produced by the free-streaming of the lighter and faster DM particles. In this case, the mass of the DM particle completely determines the suppression of the density power spectrum compared to the CDM case 
 
The computation of the halo mass function for the WDM scenario is based on the standard procedure described and tested against N-body simulations.  
The differential halo mass function (per unit $log\,M$) based on the extended Press \& Schechter approach \cite[e.g.][]{Bond1991} reads:
\begin{equation}
 {d\,\phi\over d\,logM}={1\over 6}\,{\overline{\rho}\over M}\,f(\nu)\,{d\,log\,\sigma^2\over d\,log r}\,.
\end{equation}
Here $\nu\equiv \delta_c^2(t)/\sigma^2$ depends on the linearly extrapolated density for collapse in the spherical model $\delta_c=1.686/D(t)$ and $D(t)$ is the growth factor of DM perturbations. A spherical collapse model for which $f(\nu)=\sqrt{2\nu/\pi}\,exp(-\nu/2)$ is assumed.

The key quantity entering Eq. 1 is the variance of the linear power spectrum $P(k)$ of DM perturbations (in terms of the wave-number $k=2\pi/r$). Its dependence on the spatial scale $r$ of perturbations is:
\begin{equation} 
{d\,log\,\sigma^2\over d\,log\,r}=-{1\over 2\,\pi^2\,\sigma^2(r)}\,{P(1/r)\over r^3}.
\end{equation}

In  WDM scenarios the spectrum $P_{WDM}$ is suppressed with respect to the CDM case $P_{CDM}$ below a characteristic scale depending on the mass  $m_X$ of the WDM particles. In the case of relic thermalized particles, the suppression factor can be parametrized as in Bode et al. 2001 \cite{Bode}:
\begin{equation}
T_{WDM}(k)={P_{WDM}(k)\over P_{CDM}(k)}=\Big[1+(\alpha\,k)^{2\,\mu}\Big]^{-10/\mu}\,.
\end{equation}
where $\mu=1.12$ and the quantity $\alpha$ is linked to the WDM free-streaming scale:
\begin{equation}
\alpha=0.049 \,
\Bigg[{\Omega_X\over 0.25}\Bigg]^{0.11}\,
\Bigg[{m_X\over {\rm keV}}\Bigg]^{-1.11}\,
\Bigg[{h\over 0.7}\Bigg]^{1.22}\,{h^{-1}\over \rm Mpc},  
\end{equation}
where $m_X$ is the WDM particle mass, $\Omega_X$ is the WDM density parameter ($\Omega_X$) and $h$ the Hubble constant in units of 100 km/s/Mpc.

The mass function is computed through Eq. 1 after substituting Eq. 2, with a power spectrum $P(k)=P_{WDM}(k)$ 
determined by the WDM particle mass $m_X$ after Eqs. 3 and 4. A half-mode wavenumber is defined, as the $k_{hm}$ at which the transfer function $T_{WDM}(k)$ is equal to $1/2$~\cite{Bode,Schneider}. Correspondingly, a half-mode mass $M_{hm}$ can also be defined:
\begin{equation}
\label{halfmodemass}
    M_{hm}=\frac{4\pi}{3}\rho_m \left [\pi\epsilon(2^{\mu/5}-1)^{-1/2\mu}\right]^3.
\end{equation}

\subsection{Semi-Analytic Model}
\label{subsez_soppressione_power_spectrum}
To investigate the interplay between WDM scenarios and reionization history, we use the semi-analytic model developed by Menci et al. (2018), to which we refer for further informations~\cite{Menci18}. The model retraces the collapse of dark matter halos through a Monte Carlo procedure on the basis of the merging history given by EPS formalism, at $0<z<10$~\cite{Menci18}. In this framework, the DM structures formation is determined by the power spectrum: the WDM $P(k)$ is computed by the suppression of the CDM one,  due to the particles free streaming at kpc scale, as described in the previous section.

Conversely, for sterile neutrino based cosmological scenarios, we refer to $M_{hm}$ from Lovell et al. (2020), obtained comparing CDM and WDM simulations performed within the same cosmic volume and in which the parameterization of the WDM halo mass function is given by $R_{fit}$~\cite{Lovell}:
\begin{equation}
\label{fitting}
    R_{fit}=\frac{n_{WDM}}{n_{CDM}}=\left(1+\left(\alpha\frac{M_{hm}}{M_{halo}}\right)^\beta\right)^\gamma,
\end{equation}
where $n_{CDM}$ and $n_{WDM}$ are the differential halo mass functions and $M_{halo}$ is the halo-mass. The numerical value of $\alpha$, $\beta$ and $\gamma$ coefficients changes if we consider central ($\alpha=2.3$, $\beta=0.8$, $\gamma=-1.0$) or satellite halos ($\alpha=4.2$, $\beta=2.5$, $\gamma=-0.2$)~\cite{Lovell}. 

We perform our analysis with five different sterile neutrino models, with a mass of $7.0$ keV, labelled according to the lepton asymmetry number ($L_6$), which is indicated in the last part of the name. For example, $L_6= 120$ is named LA120, $L_6= 8$ is named LA8 and so on. Among them, the models LA9, LA10 and LA11 are based on decaying-particles that are compatible with the X-ray $3.55$ keV emission line observed in galaxy clusters~\cite{Lovell}.

The semi-analytic model associates a galactic luminosity to each halo, depending on cooling process and merging history. The gas in the halo, initially set to have a density given by the universal baryon fraction and to
be at the virial temperature, cools due to atomic processes and settles into a rotationally supported disk. Then, the cooled gas is gradually converted
into stars, with a SFR given by: $\dot{M}_* = \frac{M_{gas}}{\tau_*}$, according to the Schmidt-Kennicut law with a gas conversion time scale $\tau_* = q \tau_d$, proportional to the dynamical time scale $\tau_d$ through the free parameter $q$~\cite{Menci18}. Moreover, galaxy interactions occurring in the same host halo may induce the sudden conversion of a fraction $f$ of cold gas into stars on a short time-scale given by the duration of the interaction~\cite{Menci18}. Feedback phenomena due to supernovae, AGNs and photoionization are also included, as described by Menci et al. (2018)~\cite{Menci18}. Finally, the luminosity produced by the stellar populations is computed by assuming a Salpeter IMF~\cite{Menci18}. 
In our analysis, we integrate the rest-frame UV ($\sim$1400  \AA) 
dust-corrected LF between the limits $M^{lim}_{UV}= [-25, -12]$, in order to obtain the corresponding luminosity density:
\begin{equation}
\label{densitàluminosità}
    \rho_{UV}=\int^{M^{lim}_{UV}}{\mathrm{d}M_{UV}\frac{\mathrm{d}N} {\mathrm{d}M_{UV}}R_{fit}L_{UV}},
\end{equation}
which is dominated by the contribution of systems with $M_{UV}\geq -20$ (see Section \ref{sez_reconstructiong_reionization_history}).

The number density of UV photons that actively participate to hydrogen ionization process is obtained by multiplying for two quantities~\cite{Finkelstein}: 
\begin{equation}
    \dot{N}_{ion}=f_{esc} \xi_{ion} \rho_{UV}.
\end{equation}

The ionizing photon production efficiency ($\xi_{ion}$) is expressed in $Hz/erg$ units and it describes how efficiently is possible to get UV ionizing photons from an UV continuum radiation field. Finally, the escape fraction $f_{esc}$ converts the intrinsic ionizing emissivity $\dot{N}_{ion, intrinsic}=\xi_{ion} \rho_{UV}$ into an effective one. It is defined as the fraction of ionizing photons that can escape from the source galaxy instead of being reabsorbed inside it and which therefore actively participates in the ionization of the IGM.

In our study, we model the Reionization history with different values of $f_{esc}$. Fixed escape fraction is useful to broadly characterize the Reionization history, although a universal value for $f_{esc}$ is highly unrealistic.

Nevertheless, the investigation of the degenerate quantities $f_{esc}\xi_{ion}$, which drive the Reionization process, can yield to interesting upper limits to the \mbox{escape fraction}.

Once obtained $\dot{N}_{ion}$, the equation that accounts for ionization and recombination, which regulates the evolution of the hydrogen filling fraction $Q_{HII}$ is: 
\begin{equation}
\label{eq_fillingfrac}
    \dot{Q}_{HII}=\frac{\dot{N}_{ion}}{\bar{n}_H}-\frac{Q_{HII}}{t_{rec}},
\end{equation}

where the comoving hydrogen mean density is computed as $\bar{n}_H\approx2\times10^{-7}(\Omega_b h^2/0.022)$ cm$^{-3}$ and the recombination time-scale is $t_{rec}\approx3.2$ Gyr $[(1+z)/7]^{-3}C^{-1}_{HII}$~\cite{Lapi}. We consider case B of recombination, in which electrons fallen to the ground level generate ionizing photons that are
re-absorbed by the optically thick IGM, having no consequences on the overall ionization balance. We treat the evolution of the clumping factor $C_{HII}$ with redshift, due to the effect of UVB generated by Reionization, according to, e.g., Haardt et al. 2012 ~\cite{Haardt}:
\begin{equation}
\label{eq_clump}
    C_{HII}=1+43z^{-1.71}.
\end{equation}

After the reconstruction of the Reionization history, we use the redshift evolution of the filling fraction to compute the integral:
\begin{equation}
\label{eq_taues}
    \tau_{es}(z)=c\sigma_T\bar{n}_H\int_0^z Q_{HII}(z')(1+z')^2 \left(1+\frac{\eta Y}{4X}\right)H^{-1}(z')\mathrm{d}z',
\end{equation}
in which helium is singly-ionized ($\eta=1$) at $z>4$ and doubly-ionized ($\eta=2$) at \mbox{$z<4$}. Then, the electron scattering optical depth has been compared with observational constraints on $\tau_{es}$ obtained, from CMB anisotropy, by Planck and WMAP.

 \begin{figure}[htb]
    \begin{center}
        {\includegraphics[scale=0.35]{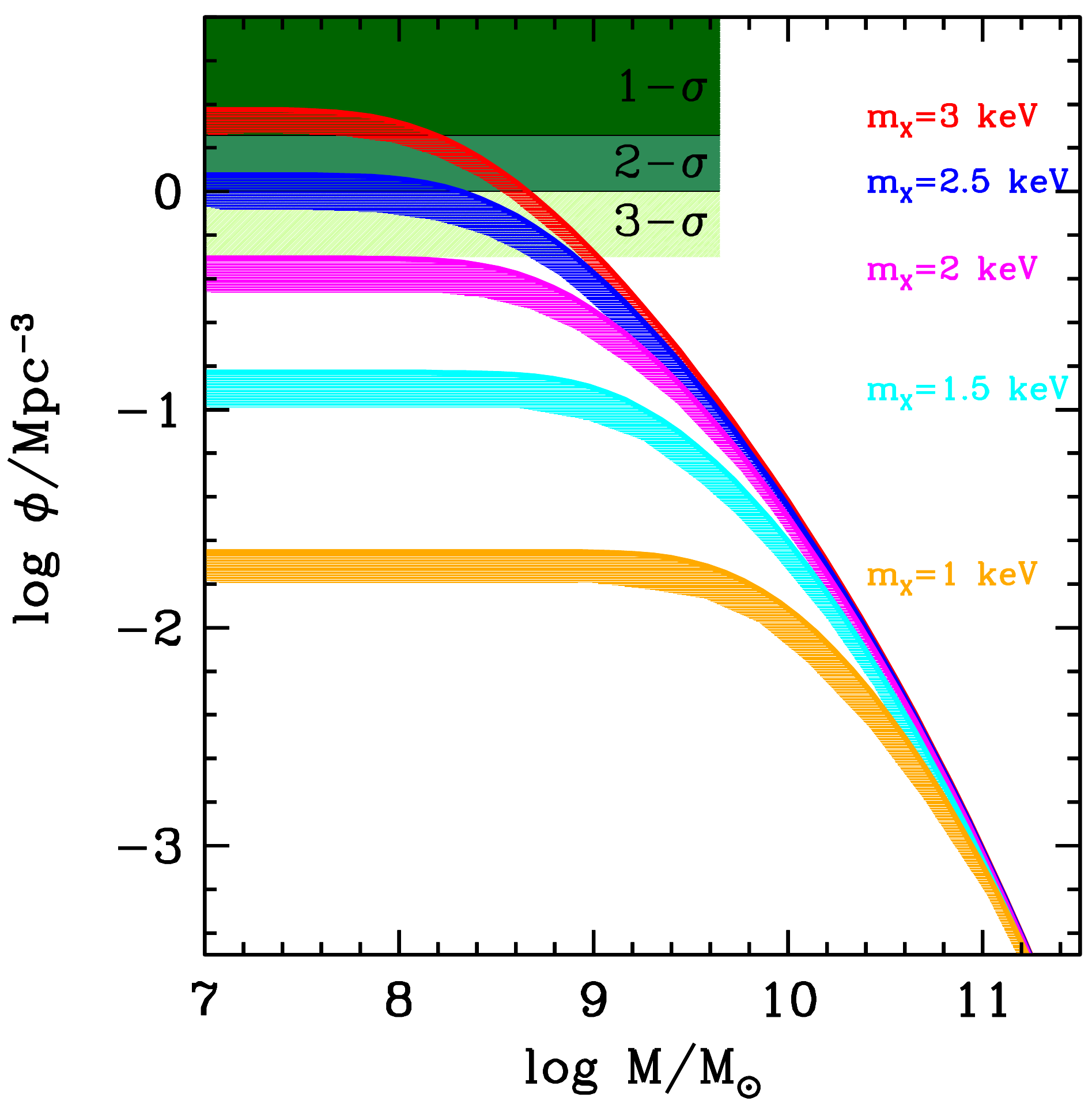}}\hspace{0.5cm}
        \caption{\it Adapted from Menci et al. 2016 \cite{Menci}: the cumulative mass functions computed at $z=6$ for different values of the WDM particle mass $m_X$ from 1 to 3 keV (bottom to top). The shaded areas correspond to the observed number density of HFF galaxies within 1-$\sigma$, 2-$\sigma$, and  3-$\sigma$  confidence levels.}
\label{fig1}
    \end{center}
\end{figure}

\section{Results}
 \subsection{Constraints on thermal WDM from the abundance of high-redshift galaxies}

We compare the halo number densities in WDM cosmologies to the observed number density ${\phi}_{obs}$ of galaxies derived by integrating the galaxy luminosity function (LF) at $z=6$ by ~\cite{Livermore} down to the faintest bin $M_{\rm UV}= -12.5$. Constraints on DM models are simply put by requiring that observed galaxies cannot outnumber their host DM halos ($\overline{\phi}\geq {\phi}_{obs}$). The reference luminosity function has been estimated from objects in the Abell~2744 and MACS~0416 cluster fields, selected on the basis of their photometric redshift.

 \begin{figure}[htb]
    \begin{center}
        {\includegraphics[scale=0.25]{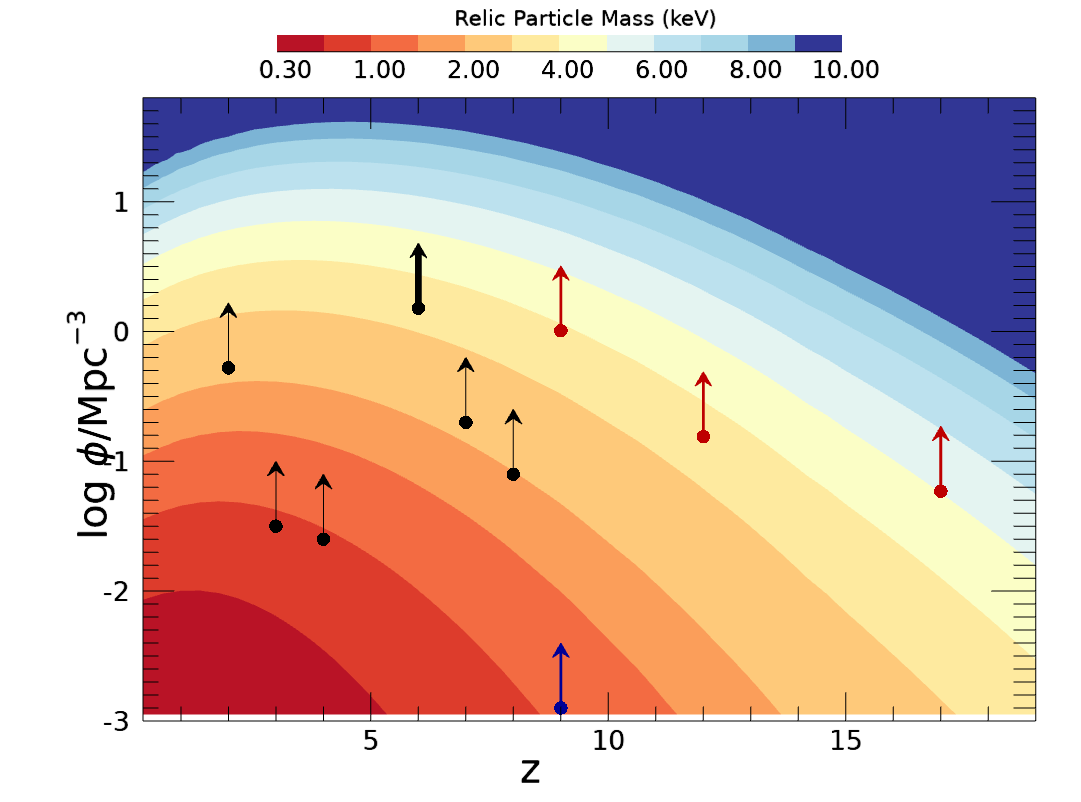}}\hspace{0.5cm}
        \caption{\it Constraints on the abundance of dark-matter halos derived from the galaxy LF at different redshifts (black arrows). The corresponding limits on the mass of WDM thermal relics are shown as colored contours. JWST measurements reaching $z>9$ lensed galaxies down to $M_{UV}\leq -12.5$ (red arrows) can significantly strengthen currents constraints if no cut-off of the LF will be found.}
\label{figJWST}
    \end{center}
\end{figure}

In Fig.~\ref{fig1} we show the cumulative mass function $\phi(>M)$ at $z=6$ for different assumed WDM particle masses. 
All the mass functions saturate to a maximum number density $\overline\phi_{m_X}\approx \phi (M_{hm})$. This is compared with the observed number density $\phi_{obs}$ of galaxies with $M_{UV}\leq -12.5$.  The condition $\phi_{obs}\leq \overline\phi_{m_X}$ yields  $m_X\gtrsim 2.9$ keV at 1-$\sigma$ level,  $m_X\geq 2.4$ keV at 2-$\sigma$ level, and $m_X\geq 2.1$ keV at 3-$\sigma$ level.
In Fig.~\ref{figJWST} we show the constraints on the thermal relic WDM particle mass from the abundance of galaxies in available observations at $z<7$, including the quoted constraint from the HFF observations at $z\sim6$, compared to what could be achieved by hypothetical JWST observations of strongly lensed galaxies at $z>9$ (red arrows). We derive the estimate by extrapolating to  $M_{UV}\leq -12.5$ the UV LF at $z=9-16$ recently published by Harikane et al. 2022 \cite{Harikane22}. If JWST will confirm a steep faint-end of the UV LF at extremely high-redshifts, thermal relic particles with masses $m_X < 5$ keV will be ruled out by galaxy abundance measurements alone.

\subsection{The reionization history in WDM scenarios}
\label{sez_reconstructiong_reionization_history}
Here we investigate the unfolding of the reionization epoch in WDM cosmologies, summarising the cited work by Romanello et al. 2021 \cite{Romanello21}.

\begin{figure}[ht]
\centering
\includegraphics[width=11 cm]{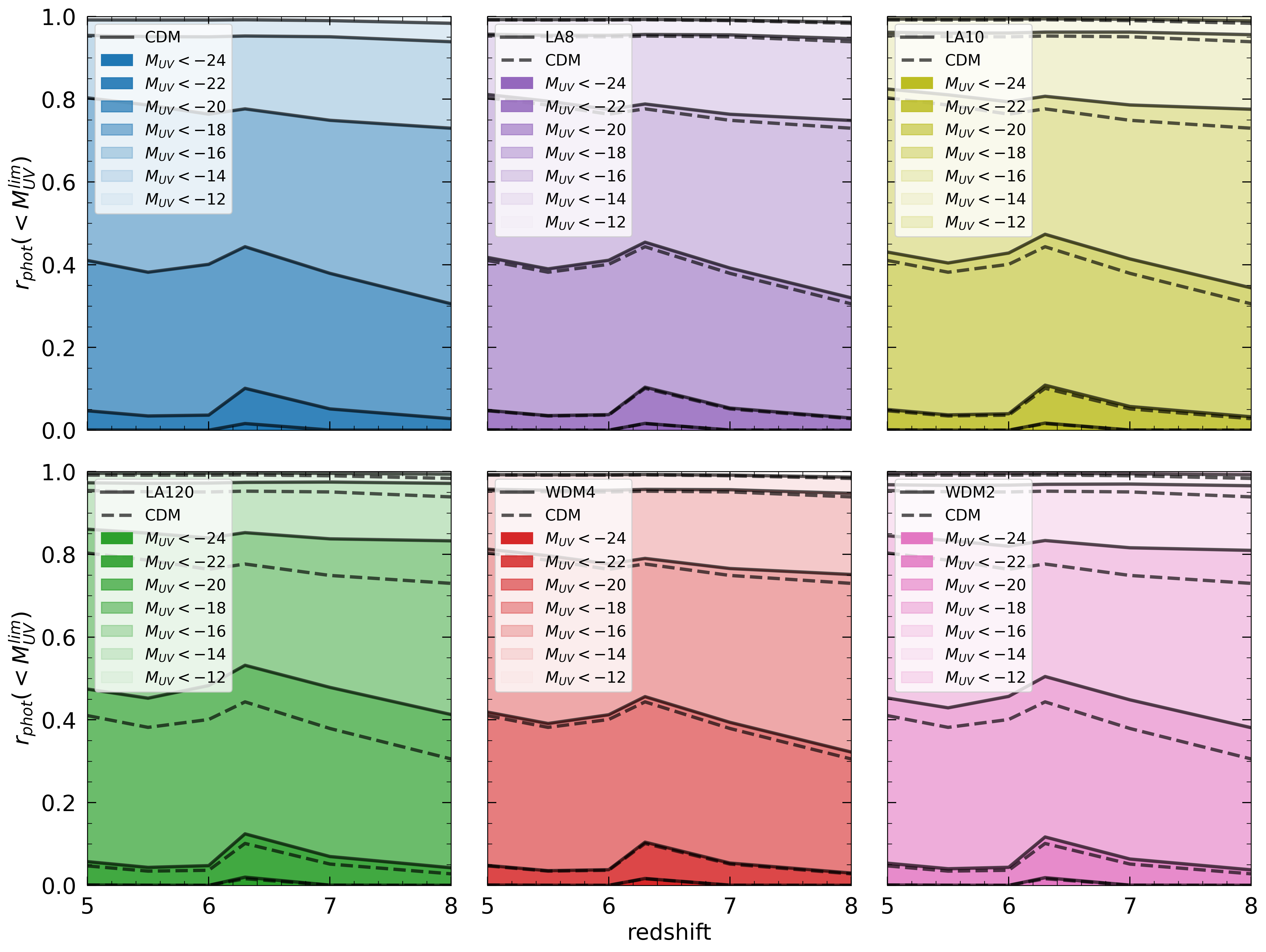}
\caption{\it  Adapted from Romanello et al. 2021 \cite{Romanello21}. The multiple panels show the integrated photons ratio $r_{phot}(<M_{lim}^{UV})$, where $\dot{N}_{ion,tot}$ is computed by integrating Equation (\ref{densitàluminosità}) between intrinsic $M_{UV}^{sup}=-12$ and $M_{UV}^{inf}=-25$. We compare with CDM two thermal WDM cosmologies (WDM3 is an intermediate case between WDM4 and WDM2), and three sterile neutrino cosmologies (here LA10 is the only representative scenario for radiatively decay Dark Matter, which is compatible with the 3.5 keV emission line observed in galaxy clusters).} \label{fig_photonsfraction}
\end{figure}

In Figure \ref{fig_photonsfraction} we plot the integrated ionizing photons ratio:
\begin{equation}
r_{phot}(<M_{lim}^{UV})=\dfrac{\dot{N}_{ion}(M_{UV}<M_{UV}^{lim})}{\dot{N}_{ion,tot}}
\end{equation}
in which we compute $\dot{N}_{ion,tot}$, using Equation (\ref{densitàluminosità}) between intrinsic $M_{UV}^{sup}=-12$ and $M_{UV}^{inf}=-25$, while the numerator is obtained by varying the upper limit of the integral from $-24$ to $-12$, including so the photons from progressively dimmer sources, until the unity is reached. 

\begin{figure}[t]
\centering
\includegraphics[width=9 cm]{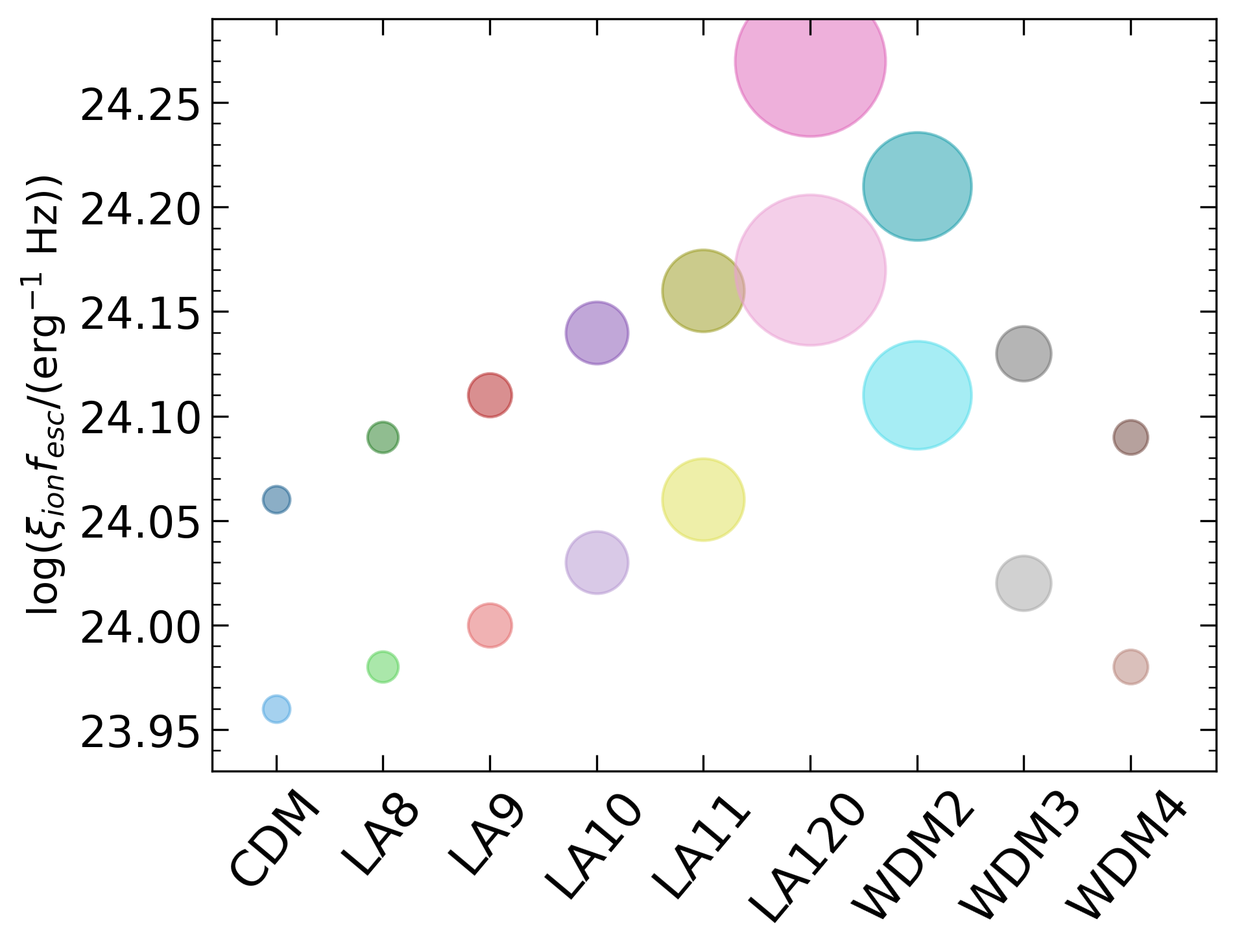}
\caption{\it Adapted from Romanello et al. 2021 \cite{Romanello21}. The product $\xi_{ion}f_{esc}$ required to ionize the IGM at $z=6.7$, for a set of different cosmologies. The dot size increases with $M_{hm}$; lighter colours refer to initial condition $Q_{HII}(z=10)=0.2$, while darker colours are for $Q_{HII}(z=9)=0.0$.}
\label{fig_degenerazione}
\end{figure}

From Figure \ref{fig_photonsfraction} we can identify two important features, through which we can understand the role and the different contribution of faint and bright galaxies during EoR. The first is the increasing of the relative contribution of the brightest systems (respectively with intrinsic $M_{UV}^{sup}<-24$, $M_{UV}^{sup}<-22$ and $M_{UV}^{sup}<-20$) with the age of universe. In the $\Lambda$CDM model, $r_{phot}(<-22)$ passes from $2.8\%$ at $z=8$, to $10\%$ at $z=6.3$. In parallel, for $M_{UV}^{sup}=-20$ we have a raise from $31\%$ at $z=8$ to $44\%$ at $z=6.3$. We can interpret this trend in the light of the hierarchical growth of cosmic structures: merging phenomena between galaxies give origin to more massive and brighter structures, increasing their overall contribution. However, the role of faint galaxies in the Reionization process is \mbox{still predominant.}

The second issue to be highlighted derives from a comparison between different cosmological scenarios, which reveals that WDM models present a relative $\dot{N}_{ion}$ higher than the CDM ones. Again, the reason resides in the effect of free-streaming, which determines a suppression in the number density of the faint-galaxies and so a decreasing in their relative contribution for each $M_{UV}^{sup}$.  The difference between cosmologies is summarized in the half-mode mass and is not negligible: if we compare CDM with LA8 and WDM4, at $z=8$ it values~$\approx$~1--2$\%$ , respectively for $M_{UV}^{sup}=-20$ and $-18$, but it increases to $8-10\%$ for WDM2 and LA120. Finally, we noted that the continue (WDM) and the dashed (CDM) lines in Figure \ref{fig_photonsfraction} approach each other with time; for example, at $z=5$ the differences between CDM and WDM2-LA120 reduce respectively to 4--6$\%$. Again, we can interpret this result by looking at the evolution of the UV LFs with $z$.

\begin{figure}[]
\centering
\includegraphics[width=10 cm]{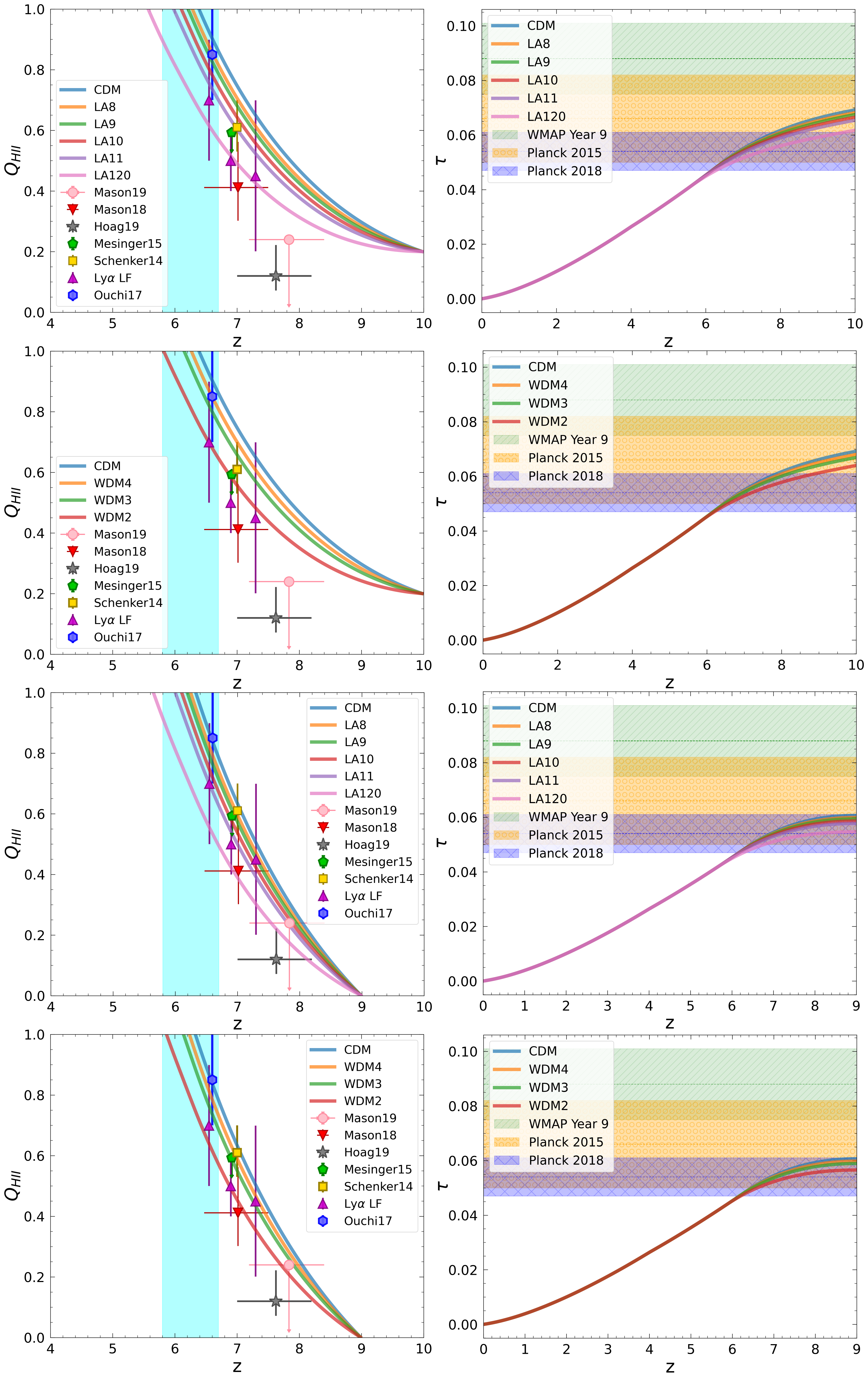}
\caption{\it Adapted from  Romanello et al. 2021 \cite{Romanello21}. \textbf{Left}: {evolution of the filling} fraction $Q_{HII}$, for sterile neutrino and thermal WDM models, with $\log(\xi_{ion}/(erg^{-1}Hz))=25.2$. The two upper panels have initial condition $Q_{HII}(z=10)=0.2$ and $f_{esc}=0.05$. The two lower panels are plotted with $Q_{HII}(z=9)=0.0$ and $f_{esc}=0.06$. The cyan shaded region indicates our fiducial late-Reionization redshift interval, $5.8<z<6.7$. The upward triangle labelled Ly$\alpha$ LF includes results by Konno et al. (2014), Konno et al. (2017) and Zheng et al. (2017)~\cite{Konno14,Konno17,Zheng17}. \textbf{Right}: electron scattering optical depth for different models, compared with measurements from Planck and WMAP~\cite{Planck15,Planck18,WMAP9}.} \label{fig_filling_tau_fixed}
\end{figure}  

The current analysis is based on intrinsic UV luminosity and it is independent from the dust extinction, which is summarized in the escape fraction value: in fact $f_{esc}$ appears only as a multiplicative constant, so it simplifies in the ratio between $\dot{N}_{ion}$. Conversely, if we consider other escape fraction dependencies, we could expect a more various behaviour.

\subsubsection{Implications on $f_{esc}$}
\label{sez_implications_fesc}

The evolution of the filling fraction with cosmic time depends also on the initial condition for Equation (\ref{eq_fillingfrac}). In particular, we choose two extreme possibilities, which are motivated both with model available in literature and with an observational point of view.

The first has $Q_{HII}(z=10)=0.2$. This assumption agrees with the $68\%$ credibility interval modelled on the marginalized distribution of the
neutral fraction ($1-Q_{HII}$), from the SFR histories and the Planck constraints on $\tau_{es}$, from Robertson et al. (2015)~\cite{Robertson}.
Similarly, it is coherent with the range of $Q_{HII}$ allowed for the model by Bouwens et al. (2015), where Reionization is complete between $z = 5.9$ and $z = 6.5$~\cite{Bouwens15}. As a second possibility, we choose $Q_{HII}(z=9)=0.0$, which is preferred by the two hydrogen neutral fraction measurements performed by Mason et al. (2019) and Hoag et al. (2018)\cite{Hoag19,Mason19}. All the others are intermediate cases.

For each of the two initial conditions we compute the number density of ionizing photons $\dot{N}_{ion}$ with different combinations of $\xi_{ion}f_{esc}$, exploring the effect of the parameters degeneracy on the reheating of IGM. Particles free-streaming has consequences on galaxy formation, determining a lack of faint-galaxies which alters the UV LF, with a general reduction in the UV luminosity density in models with a high $M_{hm}$. Thus, we obtain a delay in the IGM ionizing process, with respect to CDM.

In Figure \ref{fig_degenerazione}, we show $\log(\xi_{ion}f_{esc})$ in CDM, sterile neutrinos and thermal WDM cosmologies. Due to the great uncertainty on $f_{esc}$, we searched for the $\xi_{ion}f_{esc}$ values that ensure the completion of Reionization at $z=6.7$. We note that $\log(\xi_{ion}f_{esc})$ increases with $M_{hm}$:
a larger escape fraction and/or UV photons production efficiency are needed to complete the Reionization process in WDM scenarios. However, the quantity $\xi_{ion}$ is better constrained than $f_{esc}$, so we assume from the literature a fiducial value of $\log(\xi_{ion}/(erg^{-1}Hz))=25.2$~\cite{Finkelstein,Bouwens15}, as expected from a low metallicity single-star population. This value is coherent with the Salpeter IMF assumed in the semi-analytic model~\cite{Robertson}. We did not investigate the variation of $\xi_{ion}$ with redshift and $M_{UV}$, which we have considered negligible with respect to changes in escape fraction. Similarly, we have neglected the variation with galaxy age. These hypotheses allow us to set an upper limit to $f_{esc}$ for each different WDM particle and boundary condition. In general, models that start from $Q=0$ need a higher $f_{esc}$ value to ionize the IGM within the same $z$ range. For this reason they are more inclusive and result in a weaker constraint to the admitted escape fraction. If $f_{esc}>f_{esc}^{sup}$, Reionization process is completed outside the fiducial redshift interval.

The evolution of the  filling fraction $Q_{HII}$ in the various cosmological models and for different assumptions is summarised in Fig.~\ref{fig_filling_tau_fixed}.

\section{Conclusions}
In this paper we have first summarised the results presented in Menci et al. 2016, 2017 \cite{Menci,Menci2017} putting stringent constraints on DM models with suppressed power spectrs. The comparison of the predicted maximum number density of DM halos $\overline{\phi}$ to the observed number density ${\phi}_{obs}$ provide robust constraints through the simple condition that observed galaxies cannot outnumber their host DM halos ($\overline{\phi}\geq {\phi}_{obs}$). Remarkably, these constraints are conservative, and independent of the modeling of baryonic physics in low-mass galaxies. The mass of WDM thermal relic candidates is constrained to be $m_X\geq 2.9$ keV at 1$\sigma$ confidence level,  and $m_X\geq 2.4$ keV at $2-\sigma$ level. 
 by have been derived by comparing the maximum number density of DM halos $\overline{\phi}$ expected at redshift $z=6$ to the observed number density ${\phi}_{obs}$ of galaxies at the same redshift in the HFF.

We have then summarised the results presented in  \cite{Romanello21} on how Reionization scenarios are affected by early galaxy formation in WDM cosmologies. We have used the semi-analytic model described by Menci et al. (2018)~\cite{Menci18}, to produce the UV LF in a $\Lambda$CDM framework. We have tested some $\Lambda$WDM cosmologies, in which the contribution of the faint galaxies is suppressed: in particular, we have focused on five sterile neutrino models presented in Lovell et al. (2020)~\cite{Lovell}), and three thermal WDM models with \mbox{$m_X=$ 2--3--4 keV}. In both cases, we have found that a higher $M_{hm}$ leads to a general delay in the Reionization process. In CDM cosmology, merging between galaxies determines the rise of the intrinsic $M_{UV}<-20$ systems relative contribution to the ionizing photons budget, from $\approx$$30\%$ to $\approx$$45\%$ between $6.3<z<8$. In the WDM case, the particles free-streaming yields to a shift towards brighter sources and $r_{phot}(M_{UV}<-20)$ undergoes a further 1--10$\%$ growth, depending on cosmology; We found that in WDM cosmologies a higher $f_{esc}\xi_{ion}$ is required, in order to complete the Reionization process at the same redshift. 

The most important limits to our analysis are related to observational uncertainties. We expect significant advances thanks to improved constraints on the UV LF at very high-redshift made possible by forthcoming JWST surveys.

\printbibliography
\end{document}